\documentclass[aps,twocolumn,showpacs,footinbib]{revtex4-1}
\usepackage{mathrsfs}
\usepackage{amsfonts}
\usepackage{amssymb}
\usepackage{amsmath}
\usepackage{amscd}
\usepackage{graphicx}
\usepackage{float}
\usepackage{array}

\begin{document}

\title[Short Title]{Engineering W-type steady states for three atoms via dissipation in an optical cavity}

\author{Xin-Yu Chen$^{1}$}
\email{xychen.fzu@gmail.com}
\author{Li-Tuo Shen$^{1}$}
\author{Zhen-Biao Yang$^{2}$}
\author{Huai-Zhi Wu$^{1}$}
\author{Mei-Feng Chen$^{1}$}
\email{meifchen@126.com}

\affiliation{$^{1}$Lab of Quantum Optics, Department of Physics,
Fuzhou University, Fuzhou 350002, China\\$^{2}$Key Laboratory of
Quantum Information, University of Science and Technology of China,
CAS, Hefei 230026, China}

\begin{abstract}
We propose a scheme for dissipative preparation of W-type entangled
steady-states of three atoms trapped in an optical cavity. The
scheme is based on the competition between the decay processes into
and out of the target state. By suitable choice of system
parameters, we resolve the whole evolution process and employ the
effective operator formalism to engineer four independent decay
processes, so that the target state becomes the stationary state of
the quantum system. The scheme requires neither the preparation of
definite initial states nor precise control of system parameters and
preparation time.
\end{abstract}

 \maketitle

\section{Introduction}

Entanglement of multiple particles is not only an essential
ingredient for a test of quantum nonlocality, but also a key
resource for implementation of quantum information processing (QIP)
\cite{PRL1991-67-661,PRL1992-69-2881,PRL1993-70-1895}. Preparing
entangled states faithfully and reliably has been one of the main
tasks in quantum computation \cite{PRL2000-85-2392,cup2000}. To
achieve this, one of the main obstacles is decoherence induced by
the environment. Recently, many strategies using decoherence as a
resource have been developed in quantum computation and entanglement
engineering
\cite{PRL2011-106-090502,arXiv1110.1024v1,PRA2012-85-032111,PRA2011-84-022316,JOSAB2011-28-228,arXiv1005.2114v2,
PRA2011-83-042329,PRL2011-107-120502,PRA2010-82-054103,
PRL2011-106-020504,PRA2011-84-064302}. Schemes based on dissipative
preparation require neither the preparation of definite initial
states nor precise control of system parameters and preparation
time. Particularly, Kastoryano and Reiter \emph{et al.}
\cite{PRL2011-106-090502,arXiv1110.1024v1,PRA2012-85-032111}
proposed a scheme to produce maximally entangled states for two
atoms trapped in an optical cavity via engineering the decay
process. Busch \emph{et al.} \cite{PRA2011-84-022316} showed that
two atoms in an optical cavity can be cooled to a maximally
entangled state by employing level shifts induced by laser fields.
The distinct feature of these schemes
\cite{PRL2011-106-090502,arXiv1110.1024v1,PRA2012-85-032111,PRA2011-84-022316}
is the linear scaling of the fidelity with the cooperativity as
compared to square root scaling of the fidelity for the schemes
based on unitary dynamics. The idea of Refs.
\cite{PRL2011-106-090502,arXiv1110.1024v1,PRA2012-85-032111} has
been applied to dissipative preparation of maximally entangled
states for two atoms trapped in two coupled cavities
\cite{PRA2011-84-064302}. However, most of the previous theoretical
schemes
\cite{PRL2011-106-090502,arXiv1110.1024v1,PRA2012-85-032111,PRA2011-84-022316,JOSAB2011-28-228,arXiv1005.2114v2,
PRA2011-83-042329,PRL2011-107-120502,PRA2010-82-054103,
PRA2011-84-064302} and experiments \cite{PRL2011-107-080503}
concentrate on the preparation of entangled states of two atoms. To
our knowledge, there is no experimental report for dissipative
preparation of multipartite entangled states in cavity QED.

One of the most important multipartite entangled state is the W-type
state which has been shown to have valuable applications in QIP such
as quantum teleportation \cite{PLA2002-296-161}, quantum dense
coding \cite{PLA2003-314-267}, quantum cloning machine
\cite{PRA1998-57-2368}, etc. Recently, three-qubit W states have
been achieved in optical systems \cite{PRL2004-92-077901}, ion traps
\cite{Science2004-304-1478} and superconducting phase qubits
\cite{Nature2010-467-570}. Numerous schemes have been proposed for
generation of such states in cavity QED via unitary dynamics
\cite{PRA2000-62-062314,PRA2004-69-052314,PRA2002-65-054302,JOB2005-7-10,
PRA2006-74-024303}. However, a W state has not been realized
experimentally in cavity QED. The fidelity of schemes based on
cavity QED will suffer errors coming from spontaneous emission and
cavity decay, but these two error sources can not be decreased at
the same time in unitary dynamics \cite{PRL2011-106-090502}.

In this paper, we propose a scheme for the dissipative preparation
of W-type steady-state (the target state) of three atoms in an
optical cavity. The scheme is based on the competition between the
decays into and out of the target state. Each laser field, assisted
by the dissipative cavity mode and atomic spontaneous emission,
induces a collective atomic decay process independently. The total
decay rate between any pair of collective atomic states is the sum
of those associated with the four engineered decay processes. By
suitable choice of system parameters, the rate of decay into the
target state is much larger than that out of the target state so
that the system finally approaches the target state no matter what
the initial state is. Numerical results show that the W-type steady
entanglement can be obtained with fidelity as high as $90\%$,
despite of the cooperativity parameter C as low as $75$, where
$C=g^{2}/\kappa\gamma$.

\section{Engineering W-type Steady State}

\begin{figure}
\centering
\includegraphics[width=1\columnwidth]{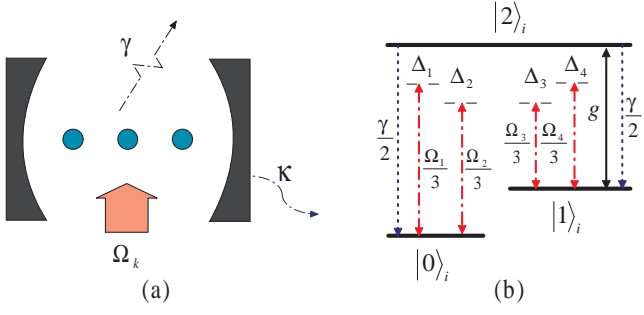} \caption{(Color
online) (a) Experimental setup for engineering W-type entangled
steady state for three atoms via dissipation in an optical cavity.
(b) Level diagram of a single atom. The
$|1\rangle\leftrightarrow|2\rangle$ transition couples resonantly
with coupling constant $g$ to the cavity field. Four off-resonance
optical lasers with detuning $\Delta_{k}$ and Rabi frequency
$\Omega_{k}$ drive the transition
$|0\rangle\leftrightarrow|2\rangle$ ($k=1,2$) and
$|1\rangle\leftrightarrow|2\rangle$ ($k=3,4$),
respectively.}\label{F1}
\end{figure}

As shown in FIG. \ref{F1}, three $\Lambda$-type atoms are trapped in
a single-mode cavity. Each atom has two ground states $|0\rangle$
and $|1\rangle$ and an excited state $|2\rangle$. The cavity mode is
coupled to the $|1\rangle\leftrightarrow|2\rangle$ transition
resonantly. Two off-resonance optical lasers, each with detuning
$\Delta_{k}$ and Rabi frequency $\Omega_{k}$, drive the transition
$|0\rangle\leftrightarrow|2\rangle$ ($k=1,2$). The
$|1\rangle\leftrightarrow|2\rangle$ transition is driven by two
other different lasers, each with detuning $\Delta_{k}$ and Rabi
frequency $\Omega_{k}$ ($k=3,4$) respectively.

It is a tough work to obtain the analytical result of the present
system for the reason that there is no interaction picture in which
the system Hamiltonian becomes time independent. When the classical
fields are sufficiently weak, the condition
$\Omega_{k}\Omega_{l}(1/\Delta_{k}+1/\Delta_{l})/2\ll|\Delta_{k}-\Delta_{l}|$
($k,l=1,2,3,4;k\neq l$) is satisfied. We can neglect the Raman
transition between two any classical fields. In this case, each
laser field, together with the dissipative cavity mode and atomic
spontaneous emission, induces a collective atomic decay process
independently. The whole dissipative process is the incoherent
combination of the four independent decay processes.

Under the rotating wave approximation, the Hamiltonian associated
with the $k$th independent decay process in the interaction picture
reads $H^{(k)}=H_{0}^{(k)}+V^{(k)}_{+}+V^{(k)}_{-}$, where
\begin{eqnarray}\label{e1}
H_{0}^{(k)}&=&\Delta_{k}
a^{\dagger}a+\sum_{m=1}^{3}\Delta_{k}|2\rangle_{m}\langle2|
\cr&&+\sum_{m=1}^{3}\Big(ga|2\rangle_{m}\langle1|+ga^{\dagger}|1\rangle_{m}\langle2|\Big),\\
\label{e2}
V^{(k)}_{+}&=&\frac{\Omega_{k}}{3}\sum_{m=1}^{3}|2\rangle_{m}\langle0|,\qquad
(k=1,2) \\ \label{e3}
V^{(k)}_{+}&=&\frac{\Omega_{k}}{3}\sum_{m=1}^{3}|2\rangle_{m}\langle1|,\qquad
(k=3,4)
\end{eqnarray}
$V^{(k)}_{-}=\big(V_{+}^{(k)}\big)^{\dagger}$, $a$ is the
annihilation operator for the cavity mode, and $g$ is the
atom-cavity coupling constant.

In the following, we assume that the system-environment interaction
is Markovian such that the evolution of the density matrix $\rho$
can be described by a master equation of Lindblad form
\begin{eqnarray}\label{e4}
\dot{\rho}&=&i\Big[\rho,H^{(k)}\Big]+\sum_{x}\bigg\{L_{x}\rho
(L_{x})^{\dagger}\cr&&-\frac{1}{2}\Big[(L_{x})^{\dagger}L_{x}\rho+\rho
(L_{x})^{\dagger}L_{x}\Big]\bigg\},
\end{eqnarray}
where the Lindblad operators $L_{x}$ represent various decay
sources. In the atom-cavity system, two decay sources will
inherently be present: spontaneous emission of the excited state to
the ground states $|0\rangle$ and $|1\rangle$ with decay rates
$\gamma_{0,m}$ and $\gamma_{1,m}$ ($m=1,2,3$), and cavity leakage at
a decay rate $\kappa$. Seven Lindblad operators associated with
dissipation can be expressed as $L_{\kappa}$ = $\sqrt{\kappa}$ $a$,
$L_{\gamma,0,m}$ = $\sqrt{\gamma_{0,m}}$ $|0\rangle_{m}\langle2|$,
$L_{\gamma,1,m}$ = $\sqrt{\gamma_{1,m}}$ $|1\rangle_{m}\langle2|$.
We assume $\gamma_{0,m=1,2,3}$ = $\gamma_{1,m=1,2,3}$ = $\gamma/2$
for simplicity.

According to Ref. \cite{PRA2012-85-032111}, based on the competition
between the unitary dynamics induced by the classical fields and the
dissipation dynamics, the time evolution of the ground subspace is
much slower than that of the excited subspace when the excited
states are not initially populated under the condition of weak
classical fields. We can adiabatically eliminate the excited cavity
field modes and excited states of the atoms. The system dynamics
will be reduced to the ground subspace in a strongly dissipative
environment. To the second order in perturbation theory, the
dynamics of the system is governed by an effective master equation
in Lindblad form
\cite{PRL2011-106-090502,arXiv1110.1024v1,PRA2012-85-032111}
\begin{eqnarray}\label{e5}
\dot{\rho}&=&i\Big[\rho,H_{eff,k}\Big]+\sum_{x}\bigg\{L^{x}_{eff,k}\rho
(L^{x}_{eff,k})^{\dagger}\cr&&-\frac{1}{2}\Big[(L^{x}_{eff,k})^{\dagger}L^{x}_{eff,k}\rho+\rho
(L^{x}_{eff,k})^{\dagger}L^{x}_{eff,k}\Big]\bigg\}.
\end{eqnarray}
It contains the effective Hamiltonian $H_{eff,k}$ and the effective
Lindblad operator $L_{eff,k}^{x}$ ($L_{x}$ represents seven Lindblad
operators defined previously)
\begin{eqnarray}\label{e6}
H_{eff,k}&=&-\frac{1}{2}V_{-}\Big(H_{NH,k}^{-1}+(H_{NH,k}^{-1})^{\dagger}\Big)V_{+},\\
\label{e7} L_{eff,k}^{x}&=&L_{x}H_{NH,k}^{-1}V_{+},
\end{eqnarray}
where $H^{-1}_{NH,k}$ is the inverse of the non-Hermitian
Hamiltonian
$H_{NH,k}=H_{0}^{(k)}-\frac{i}{2}\sum_{x}(L_{x})^{\dag}L_{x}$.

It will be convenient to work in the Fourier transformed basis of
the atomic ground states: $\{|000\rangle, |S_{1,j}\rangle,
|S_{2,j}\rangle, |111\rangle\}$ $(j=1,2,3)$, where
\begin{eqnarray}\label{e8}
|S_{1,j}\rangle&=&\frac{1}{\sqrt{3}}(e^{i\frac{2j\pi}{3}}|100\rangle
+e^{i\frac{4j\pi}{3}}|010\rangle+|001\rangle),\cr
|S_{2,j}\rangle&=&\frac{1}{\sqrt{3}}(e^{i\frac{2j\pi}{3}}|110\rangle
+e^{i\frac{4j\pi}{3}}|101\rangle+|011\rangle),
\end{eqnarray}
$|S_{1,3}\rangle=\frac{1}{\sqrt{3}}(|100\rangle
+|010\rangle+|001\rangle)$ is the desired W-type entangled state.

\begin{table*}[ht]
{\bf \caption{The decay rates $\mu_{eff,|y\rangle,|S_{1,3}\rangle}$
and $\mu_{eff,|S_{1,3}\rangle,|y\rangle}$ correspond to the
effective decay channels from $|y\rangle$ to $|S_{1,3}\rangle$ and
from $|S_{1,3}\rangle$ to $|y\rangle$ respectively by combining four
independent decay processes in Eq. (\ref{e9}).}}\label{t1}
\begin{center}
\begin{tabular}{lp{3in}lp{3in}}
\hline  $|y\rangle$&$\mu_{eff,|y\rangle,|S_{1,3}\rangle}$ &
$\mu_{eff,|S_{1,3}\rangle,|y\rangle}$
\\
\hline
$|000\rangle$&$\sum\limits_{k=1}^2\Big(\Big|\frac{\sqrt{\kappa}
g\Omega_{k}}{\sqrt{3}\tilde{R}_{1,k}}\Big|^2+3\Big|\frac{\sqrt{\gamma}
\tilde{\delta}_{k}\Omega_{k}}{3\sqrt{6}\tilde{R}_{1,k}}\Big|^2\Big)$
&$\sum\limits_{k=3}^{4}3\Big|\frac{\sqrt{\gamma}
\tilde{\delta}_{k}\Omega_{k}}{3\sqrt{6}\tilde{R}_{1,k}}\Big|^2$
\\
$|S_{1,1}\rangle$,
$|S_{1,2}\rangle$&$\sum\limits_{k=1}^{2}3\Big|\frac{\sqrt{\gamma}
\tilde{\delta}_{k}\Omega_{k}}{18\sqrt{2}\tilde{R}_{2,k}}-i\frac{\sqrt{\gamma}
\Omega_{k}}{6\sqrt{2}\tilde{\Delta}_{k}}\Big|^2+\sum\limits_{k=3}^{4}3\Big|\frac{\sqrt{\gamma}
\tilde{\delta}_{k}\Omega_{k}}{9\sqrt{2}\tilde{R}_{1,k}}\Big|^2$&$\sum\limits_{k=1}^{2}3\Big|\frac{\sqrt{\gamma}
\tilde{\delta}_{k}\Omega_{k}}{9\sqrt{2}\tilde{R}_{2,k}}\Big|^2+\sum\limits_{k=3}^{4}3\Big|\frac{\gamma
\tilde{\delta}_{k}\Omega_{k}}{9\sqrt{2}\tilde{R}_{1,k}}\Big|^2$
\\
$|S_{2,1}\rangle$,
$|S_{2,2}\rangle$&$\sum\limits_{k=3}^{4}3\Big|\frac{\sqrt{\gamma}
\tilde{\delta}_{k}\Omega_{k}}{9\sqrt{2}\tilde{R}_{2,k}}\Big|^2$&$\sum\limits_{k=1}^{2}3\Big|\frac{\sqrt{\gamma}
\tilde{\delta}_{k}\Omega_{k}}{9\sqrt{2}\tilde{R}_{2,k}}\Big|^2$
\\
$|S_{2,3}\rangle$&$\sum\limits_{k=3}^{4}3\Big|\frac{2\sqrt{\gamma}
\tilde{\delta}_{k}\Omega_{k}}{9\sqrt{2}\tilde{R}_{2,k}}\Big|^2$&$\sum\limits_{k=1}^{2}\Big(\Big|\frac{2\sqrt{\kappa}
g\Omega_{k}}{3\tilde{R}_{2,k}}\Big|^2+3\Big|\frac{2\sqrt{\gamma}
\tilde{\delta}_{k}\Omega_{k}}{9\sqrt{2}\tilde{R}_{2,k}}\Big|^2\Big)$
\\
\hline
\end{tabular}
\end{center}
\end{table*}

\begin{table}[h]
{\bf \caption{The decay rates $\mu_{eff,|y\rangle,|111\rangle}$ and
$\mu_{eff,|111\rangle,|y\rangle}$ correspond to the effective decay
channels from $|y\rangle$ to $|111\rangle$ and from $|111\rangle$ to
$|y\rangle$ respectively by combining four independent decay
processes in Eq. (\ref{e9}).}}\label{t2}
\begin{center}
\begin{tabular}{lp{1in}lp{2in}}
\hline $|y\rangle$&$\mu_{eff,|y\rangle,|111\rangle}$
&$\mu_{eff,|111\rangle,|y\rangle}$
\\
\hline $|S_{2,1}\rangle$,
$|S_{2,2}\rangle$&$\sum\limits_{k=3}^{4}3\Big|\frac{\sqrt{\gamma}
\tilde{\delta}_{k}\Omega_{k}}{3\sqrt{6}\tilde{R}_{3,k}}\Big|^2$&$\sum\limits_{k=1}^{2}3\Big|\frac{\sqrt{\gamma}
\Omega_{k}}{3\sqrt{6}\tilde{\Delta}_{k}^{2}}\Big|^2$
\\
$|S_{2,3}\rangle$&$\sum\limits_{k=3}^{4}3\Big|\frac{\sqrt{\gamma}
\tilde{\delta}_{k}\Omega_{k}}{3\sqrt{6}\tilde{R}_{3,k}}\Big|^2$&$\sum\limits_{k=1}^{2}\Big(\Big|\frac{\sqrt{\kappa}
g\Omega_{k}}{\sqrt{3}\tilde{R}_{3,k}}\Big|^2+3\Big|\frac{\sqrt{\gamma}
\tilde{\delta}_{k}\Omega_{k}}{3\sqrt{6}\tilde{R}_{3,k}}\Big|^2\Big)$
\\
\hline
\end{tabular}
\end{center}
\end{table}

Applying Eq. (\ref{e6}) and (\ref{e7}) to each decay process, we
derive the corresponding effective Hamiltonian and effective
Lindblad operators due to the competition between unitary and
dissipative dynamics (see Appendix A). We insert these effective
operators back into Eq. (\ref{e5}) and use the projection operator
approach for the density operator (i.e. $\langle
y|\dot{\rho}|y\rangle$). We note that $\langle
y|\dot{\rho}|y\rangle$ contains only diagonal terms , which
indicates there are no coherence between the states. Hence, we can
reduce the effective master equation Eq.(\ref{e5}) to the rate
equation for the population of states only. We also notice that the
effective Hamiltonian does not induce any transition between the
transformed basis states. Therefore, the transitions between these
basis states are caused by collective decays. We obtain the
effective decay rate
\begin{eqnarray}\label{e9}
\mu_{eff,|y\rangle,|z\rangle}=\sum_{k=1}^{4}\mu_{eff,|y\rangle,|z\rangle}^{(k)}
\end{eqnarray}
from the basis state $|y\rangle$ to $|z\rangle$ by combining four
decay processes. To assure that the state $|S_{1,3}\rangle$ becomes
the stationary state of the atom-cavity system, we need to choose
detuning $\Delta_{k}$ such that the transition out of
$|S_{1,3}\rangle$ can be strongly suppressed, while almost all the
population of the undesired states are driven into the target state
by collective decay. Then we keep the terms with respect to
$|S_{1,3}\rangle$. These terms are summarized in Table 1, where
$\tilde{\Delta}_{k}=\Delta_{k}-\frac{i\gamma}{2}$,
$\tilde{\delta}_{k}=\Delta_{k}-\frac{i\kappa}{2}$,
$\tilde{R}_{n,k}=\tilde{\Delta}_{k}\tilde{\delta}_{k}-ng^{2}$. In
particular, $|111\rangle$ can not be driven into the target state
directly. Effective decay from $|111\rangle$ to $|S_{1,3}\rangle$ is
mediated by \{$|S_{2,1}\rangle$, $|S_{2,2}\rangle$,
$|S_{2,3}\rangle$\}. Table 2 shows the decay rates between
$|111\rangle$ and \{$|S_{2,1}\rangle$, $|S_{2,2}\rangle$,
$|S_{2,3}\rangle$\}. Under the condition $g$ $\gg$ $\Omega_{k}$,
$\kappa$, $\gamma$, if we set $\Delta_{1}=0$, $\Delta_{2}=g$,
$\Delta_{3}=\sqrt{3}g$, $\Delta_{4}=\sqrt{2}g$, then the conditions
$\mu_{eff,|y\rangle,|S_{1,3}\rangle}\gg\mu_{eff,|S_{1,3}\rangle,|y\rangle}$
and
$\mu_{eff,|y\rangle,|111\rangle}\sim\mu_{eff,|111\rangle,|y\rangle}$
are satisfied. Thus, we can obtain state $|S_{1,3}\rangle$ with high
fidelity starting from random initial states, where the stationary
state fidelity $F=|\langle S_{1,3}|\rho|S_{1,3}\rangle|=P_{1,3}$.

\begin{figure}
\centering {\label{F2a}
\includegraphics[width=1\columnwidth]{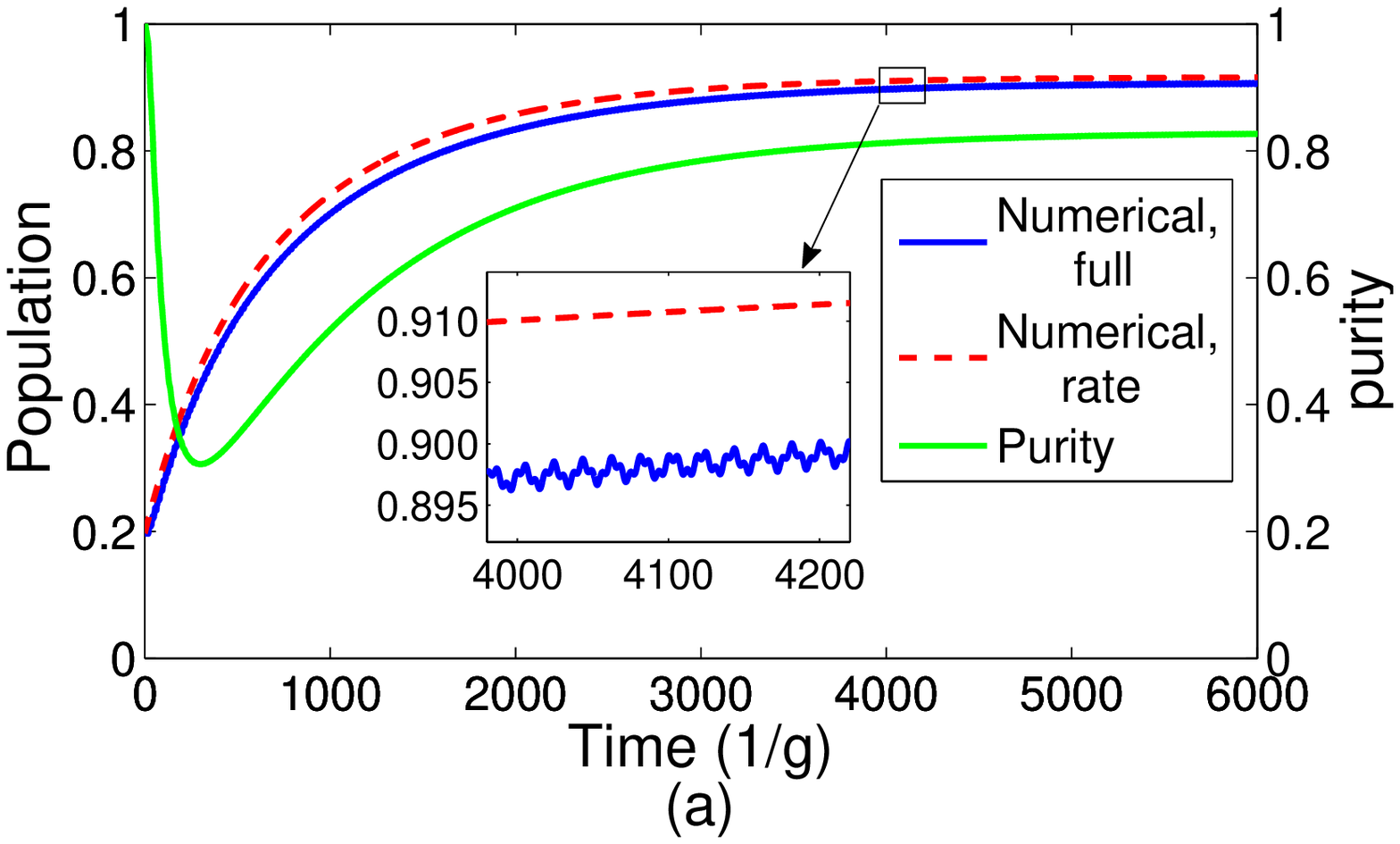}}
{\label{F2b}
\includegraphics[width=1\columnwidth]{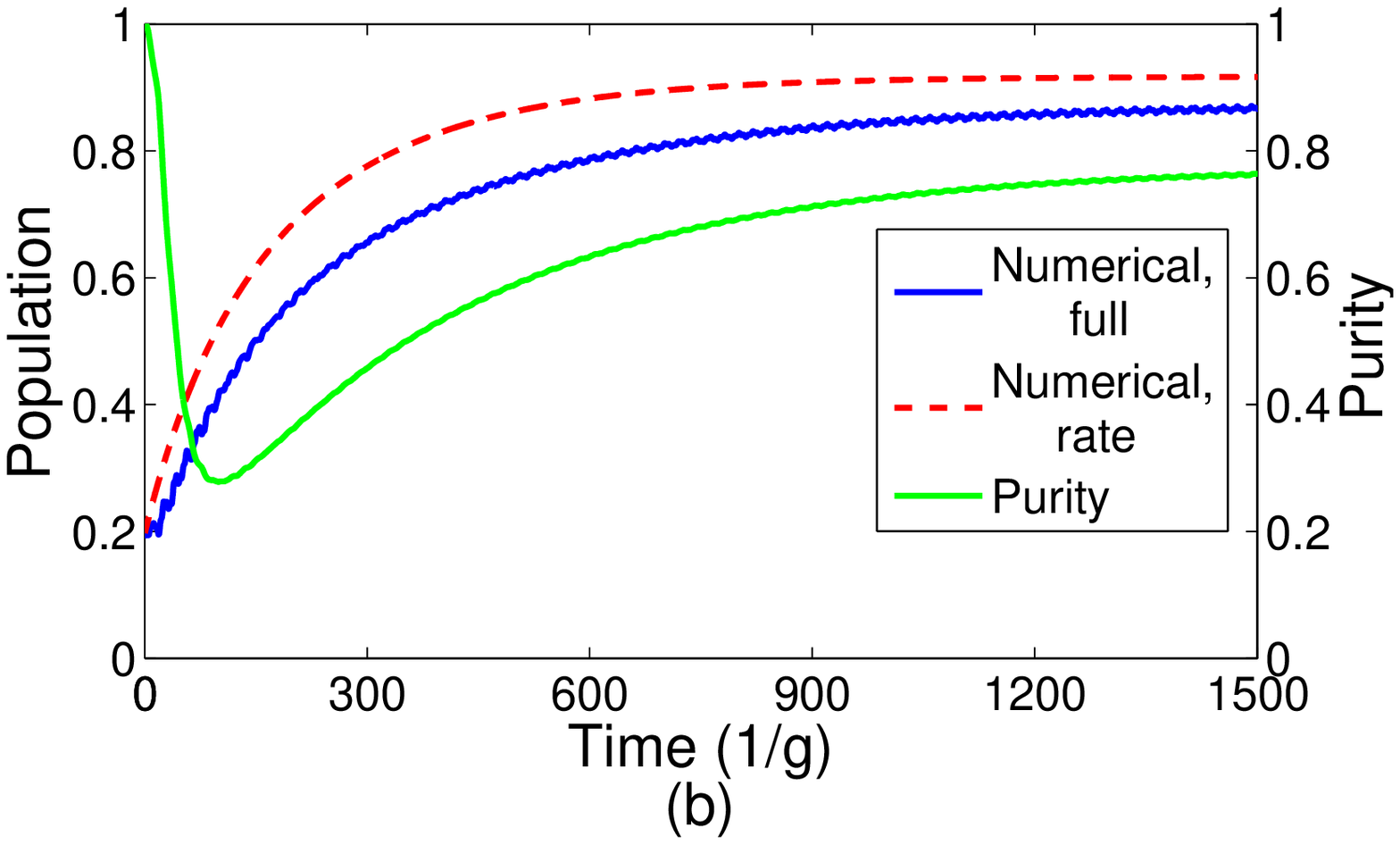}}
\caption{(Color online) The populations of the target state
$|S_{1,3}\rangle$ (left axis) and the purity of the system (green
solid line, right axis) as a function of time for a random initial
state. The curves are plotted for a set of optimal parameters
$C=80$, $\gamma=1.5\kappa$, $\Omega_{1}=\Omega_{3}=\Omega$,
$\Omega_{4}=2\Omega_{2}=1.2\Omega$. (a) $\Omega=0.04g$. (b)
$\Omega=0.08g$. Numerical results in Eq. (\ref{e10}) (red, short
dash) correspond with numerical curves obtained from the full master
equation (blue, solid line) well.}\label{F2}
\end{figure}

In order to evaluate the performance of the scheme, we insert the
decay rates obtained from the effective operators, and get the rate
equations
\begin{eqnarray}\label{e10}
\dot{P}_{y}=\sum_{z\neq
y}\big(\mu_{eff,|y\rangle,|z\rangle}P_{z}-\mu_{eff,|z\rangle,|y\rangle}P_{y}\big),
\end{eqnarray}
where $P_{y}$ is the population of state $|y\rangle$. Numerical
solution in FIG. 2(a) illustrates that we can obtain state
$|S_{1,3}\rangle$ with high fidelity above $91\%$ and the purity
$\eta$ above $82\%$, where $\eta=Tr(\rho^{2})$, in a time $6000/g$.
FIG. \ref{F2} also shows numerical results in Eq. (\ref{e10})
correspond well with numerical curves obtained from the full master
equation. The discrepancy can be attributed to the fact that we
neglect the Raman transition between any two classical fields.

Now let's consider how the Rabi frequency $\Omega$ affects the
fidelity of the steady state and the convergence speed of our
scheme. The convergence speed is primarily governed by the
magnitudes of $\Omega$. From FIG. \ref{F2}, we notice that the
convergence speed for the bigger $\Omega$ is about four times larger
than that for the smaller one. This is because that the decay rate
is proportional to the square of the Rabi frequency. However, the
Rabi frequency should not exceed a certain amount, otherwise the
condition of weak classical fields will break down, and the Raman
transitions between any two classical fields and the populations of
the excited states should be taken into consideration. Thus, as long
as the conditions
$\Omega_{k}\Omega_{l}(1/\Delta_{k}+1/\Delta_{l})/2\ll|\Delta_{k}-\Delta_{l}|$
and $g\gg\Omega$ are satisfied, an appropriate increase of $\Omega$
can speed up the convergence greatly, but it will decrease the
fidelity slightly.

\begin{figure}
\centering
\includegraphics[width=1\columnwidth]{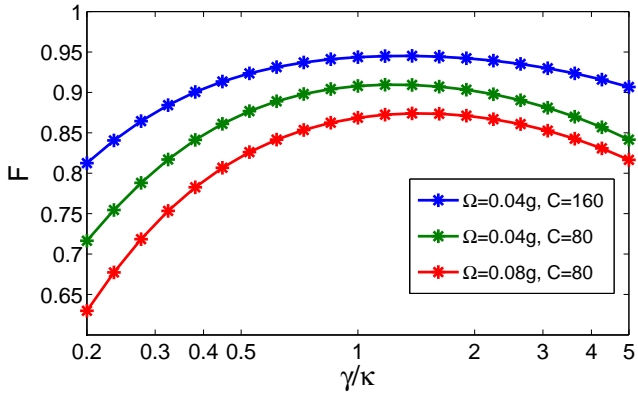} \caption{(Color
online) Stationary state fidelity $F$ as a function of
$\gamma/\kappa$ for different values of $C$ and $\Omega$.}\label{F3}
\end{figure}

\begin{figure}
\centering
\includegraphics[width=1\columnwidth]{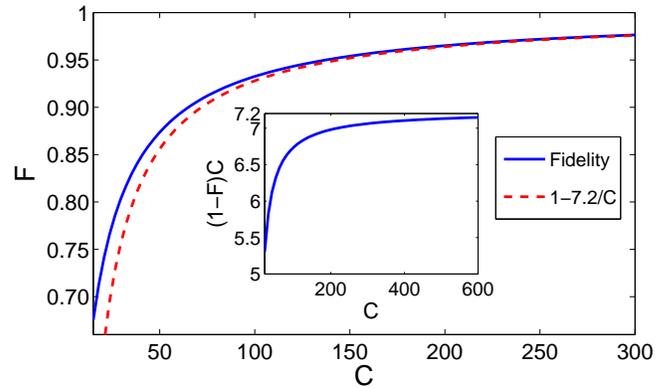} \caption{(Color
online) Stationary state fidelity $F$ as a function of cooperativity
parameter $C$ for $\gamma=1.5\kappa$, $\Omega=0.04g$. The inset
gives the coefficient of the linear scaling in $F$ as a function of
$C$.}\label{F4}
\end{figure}

FIG. \ref{F3} shows the influence of $\gamma/\kappa$ on the fidelity
$F$ of the stationary state for different values of $C$ and
$\Omega$. We find that our scheme works best when
$\gamma\in[0.8\kappa,1.8\kappa]$. We set $\gamma=1.5\kappa$ in this
paper. FIG. \ref{F4} shows the stationary state fidelity $F$ as a
function of the cooperativity parameter $C$ for $\Omega=0.04g$. Then
we carry out the curve fitting for the numerical results of Eq.
(\ref{e10}) with the least square method, and obtain the error
scaling as $1-F\propto C^{-1}$ which is similar with the scaling for
schemes of Refs.
\cite{PRL2011-106-090502,arXiv1110.1024v1,PRA2012-85-032111}. From
the inset of FIG. \ref{F4}, we find out the actual constants for
maximizing the fidelity as follows
\begin{eqnarray}\label{e11}
1-F\approx7.2C^{-1}.
\end{eqnarray}
Fidelity $F$ sees a dramatic increase as the cooperativity parameter
$C$ augments in FIG. \ref{F4}. The fidelity over $90\%$ is
attainable under a cooperativity parameter $C$ as low as $75$.

Nowadays the experimental parameters ($g$, $\gamma/2$, $\kappa/2$)
$/2\pi\approx$($34$, $2.5$, $4.1$)MHz and $C\approx28$ are
achievable
\cite{JPB2005-38-S551,PRL2006-97-083602,PRL2008-101-203602}. Then
the W-type steady states with the fidelity above $75\%$ can be
obtained, roughly in a time $5000/g\approx23\mu s$. Compared to
schemes based on unitary dynamics in cavity QED whose optimal result
is $1-F\propto C^{-1/2}$ \cite{PRL2003-91-097905}, the linear
scaling of $F$ in the present scheme has an improvement on the
cooperativity parameter. Therefore, the proposed scheme is very
promising to be realized based on the present QED techniques, and
the idea can also be generalized to other systems.

\section{Conclusion}

In conclusion, we have proposed a scheme for dissipative preparation
of W-type entangled steady states of three $\Lambda$-atoms in a
single mode optical cavity by engineering the effective decay
processes. The dissipative dynamics induced by the external fields
and dissipative cavity mode leads to the competition between decays
into and out of the target state. By suitable choice of the
parameters, the former can dominate the latter so that the target
state is the steady state of the system. We have shown that a W
state with a high fidelity can be obtained with presently available
cooperativity.

\section{Acknowledgement}

X.Y.C., L.T.S., H.Z.W., and C.M.F. acknowledge support from the
National Fundamental Research Program under Grant No. 2012CB921601,
National Natural Science Foundation of China under Grant No.
10974028, the Doctoral Foundation of the Ministry of Education of
China under Grant No. 20093514110009, and the Natural Science
Foundation of Fujian Province under Grant No. 2009J06002. Z.B.Y. is
supported by the National Basic Research Program of China under
Grants No. 2011CB921200 and No. 2011CBA00200, and the China
Postdoctoral Science Foundation under Grant No. 20110490828.

\appendix

\section{Derivation of the effective decay process}

We now consider the $1$st independent decay process induced by the
$1$st optical laser with Rabi frequencies $\Omega_{1}$ and detuning
$\Delta_{1}$, driving independently the transition
$|0\rangle\leftrightarrow|2\rangle$, as shown in Fig. \ref{F1}(b).
In a rotating frame, the Hamiltonian of this system in the
interaction picture reads
$H^{(1)}=H_{0}^{(1)}+V^{(1)}_{+}+V^{(1)}_{-}$, where
\begin{eqnarray}\label{a1}
H_{0}^{(1)}&=&\Delta_{1}
a^{\dagger}a+\sum_{m=1}^{3}\Delta_{1}|2\rangle_{m}\langle2|\cr&&
+\sum_{m=1}^{3}\Big(ga|2\rangle_{m}\langle1|+ga^{\dagger}|1\rangle_{m}\langle2|\Big),\\
\label{a2}
V^{(1)}_{+}&=&\frac{\Omega_{1}}{3}\sum_{m=1}^{3}|2\rangle_{m}\langle0|.
\end{eqnarray}

Under the condition of weak classical laser fields, we can
adiabatically eliminate the excited cavity field modes and excited
states of the atoms when the excited states are not initially
populated. Applying Eq. (\ref{e6}) and (\ref{e7}) to our setup, we
obtain the effective Hamiltonian and Lindblad operators
\begin{widetext}
\begin{eqnarray}\label{a3}
H_{eff,1}&=&\textrm{Re}\bigg[\frac{
\tilde{\delta}_{1}\Omega_{1}^{2}}{3\tilde{R}_{1,1}}\bigg]|000\rangle\langle000|+\textrm{Re}\bigg[\frac{
\tilde{\delta}_{1}\Omega_{1}^{2}}{18\tilde{R}_{2,1}}\bigg](|S_{1,1}\rangle\langle
S_{1,1}|+|S_{1,2}\rangle\langle S_{1,2}|+4|S_{1,3}\rangle\langle
S_{1,3}|) \cr&&+\textrm{Re}\bigg[\frac{
\Omega_{1}^{2}}{6\tilde{\Delta}_{1}}\bigg](|S_{1,1}\rangle\langle
S_{1,1}|+|S_{1,2}\rangle\langle S_{1,2}|)
\cr&&+\textrm{Re}\bigg[\frac{
\Omega_{1}^{2}}{9\tilde{\Delta}_{1}}\bigg](|S_{2,1}\rangle\langle
S_{2,1}|+|S_{2,2}\rangle\langle S_{2,2}|)+\textrm{Re}\bigg[\frac{
\tilde{\delta}_{1}\Omega_{1}^{2}}{9\tilde{R}_{3,1}}\bigg]|S_{2,3}\rangle\langle
S_{2,3}|,\\
\label{a4} L_{eff,1}^{\kappa}&=&-\frac{\sqrt{\kappa}
g\Omega_{1}}{\sqrt{3}\tilde{R}_{1,1}}|S_{1,3}\rangle\langle000|-\frac{\sqrt{\kappa}
g\Omega_{1}}{\sqrt{3}\tilde{R}_{3,1}}|111\rangle\langle S_{2,3}|
\cr&&-\frac{\sqrt{\kappa}
g\Omega_{1}}{3\tilde{R}_{2,1}}(-e^{-i\frac{2\pi}{3}}|S_{2,1}\rangle\langle
S_{1,2}|-e^{i\frac{2\pi}{3}}|S_{2,2}\rangle\langle
S_{1,1}|+2|S_{2,3}\rangle\langle S_{1,3}|),\\ \label{a5}
L_{eff,1}^{\gamma,0,m}&=&\frac{\sqrt{\gamma}
\tilde{\delta}_{1}\Omega_{1}}{3\sqrt{2}\tilde{R}_{1,1}}|000\rangle\langle000|
\cr&&+\Big(\frac{\sqrt{\gamma}
\tilde{\delta}_{1}\Omega_{1}}{18\sqrt{2}\tilde{R}_{2,1}}+i\frac{\sqrt{\gamma}
\Omega_{1}}{6\sqrt{2}\tilde{\Delta}_{1}}\Big)(|S_{1,1}\rangle\langle
S_{1,1}|+|S_{1,2}\rangle\langle S_{1,2}|)
\cr&&+\Big(\frac{\sqrt{\gamma}
\tilde{\delta}_{1}\Omega_{1}}{18\sqrt{2}\tilde{R}_{2,1}}-i\frac{\sqrt{\gamma}
\Omega_{1}}{6\sqrt{2}\tilde{\Delta}_{1}}\Big)(e^{i\frac{2m\pi}{3}}|S_{1,3}\rangle\langle
S_{1,1}|+e^{-i\frac{2m\pi}{3}}|S_{1,3}\rangle\langle S_{1,2}|)
\cr&&-\frac{\sqrt{\gamma}
\tilde{\delta}_{1}\Omega_{1}}{18\sqrt{2}\tilde{R}_{2,1}}(2e^{i\frac{2m\pi}{3}}|S_{1,1}\rangle\langle
S_{1,2}|+2e^{-i\frac{2m\pi}{3}}|S_{1,2}\rangle\langle S_{1,1}|)
\cr&&+\frac{\sqrt{\gamma}
\tilde{\delta}_{1}\Omega_{1}}{9\sqrt{2}\tilde{R}_{2,1}}(-e^{-i\frac{2m\pi}{3}}|S_{1,1}\rangle\langle
S_{1,3}|-e^{i\frac{2m\pi}{3}}|S_{1,2}\rangle\langle
S_{1,3}|+2|S_{1,3}\rangle\langle S_{1,3}|) \cr&&+\frac{\sqrt{\gamma}
\tilde{\delta}_{1}\Omega_{1}}{9\sqrt{2}\tilde{R}_{3,1}}(e^{i\frac{2(m-1)\pi}{3}}|S_{2,1}\rangle\langle
S_{2,3}|+e^{-i\frac{2(m-1)\pi}{3}}|S_{2,2}\rangle\langle
S_{2,3}|+|S_{2,3}\rangle\langle S_{2,3}|) \cr&&+\frac{\sqrt{\gamma}
\Omega_{1}}{9\sqrt{2}\tilde{\Delta}_{1}}(e^{-i\frac{2(m-1)\pi}{3}}|S_{2,1}\rangle\langle
S_{2,1}|+|S_{2,2}\rangle\langle
S_{2,1}|+e^{i\frac{2(m-1)\pi}{3}}|S_{2,3}\rangle\langle
S_{2,1}|)\cr&&+\frac{\sqrt{\gamma}
\Omega_{1}}{9\sqrt{2}\tilde{\Delta}_{1}}(|S_{2,1}\rangle\langle
S_{2,2}|+e^{i\frac{2(m-1)\pi}{3}}|S_{2,2}\rangle\langle
S_{2,2}|+e^{-i\frac{2(m-1)\pi}{3}}|S_{2,3}\rangle\langle
S_{2,2}|),\\ \label{a6} L_{eff,1}^{\gamma,1,m}&=&\frac{\sqrt{\gamma}
\tilde{\delta}_{1}\Omega_{1}}{3\sqrt{6}\tilde{R}_{1,1}}(e^{-i\frac{2m\pi}{3}}|S_{1,1}\rangle\langle000|
+e^{i\frac{2m\pi}{3}}|S_{1,2}\rangle\langle000|+|S_{1,3}\rangle\langle000|)
\cr&&+\Big(\frac{\sqrt{\gamma}
\tilde{\delta}_{1}\Omega_{1}}{18\sqrt{2}\tilde{R}_{2,1}}+i\frac{\sqrt{\gamma}
\Omega_{1}}{6\sqrt{2}\tilde{\Delta}_{1}}\Big)(e^{-i\frac{2(m-2)\pi}{3}}|S_{2,1}\rangle\langle
S_{1,1}|+e^{i\frac{2(m-2)\pi}{3}}|S_{2,2}\rangle\langle S_{1,2}|)
\cr&&+\Big(\frac{\sqrt{\gamma}
\tilde{\delta}_{1}\Omega_{1}}{18\sqrt{2}\tilde{R}_{2,1}}-i\frac{\sqrt{\gamma}
\Omega_{1}}{6\sqrt{2}\tilde{\Delta}_{1}}\Big)(e^{i\frac{2m\pi}{3}}|S_{2,3}\rangle\langle
S_{1,1}|+e^{-i\frac{2m\pi}{3}}|S_{2,3}\rangle\langle S_{1,2}|)
\cr&&-\frac{\sqrt{\gamma}
\tilde{\delta}_{1}\Omega_{1}}{9\sqrt{2}\tilde{R}_{2,1}}(e^{i\frac{2\pi}{3}}|S_{2,2}\rangle\langle
S_{1,1}|+e^{-i\frac{2\pi}{3}}|S_{2,1}\rangle\langle S_{1,2}|)
\cr&&+\frac{\sqrt{\gamma}
\tilde{\delta}_{1}\Omega_{1}}{9\sqrt{2}\tilde{R}_{2,1}}(-e^{i\frac{2(m-1)\pi}{3}}|S_{2,1}\rangle\langle
S_{1,3}|-e^{-i\frac{2(m-1)\pi}{3}}|S_{2,2}\rangle\langle
S_{1,3}|+2|S_{2,3}\rangle\langle S_{1,3}|) \cr&&+\frac{\sqrt{\gamma}
\Omega_{1}}{3\sqrt{6}\tilde{\Delta}_{1}}(e^{-i\frac{2m\pi}{3}}|111\rangle\langle
S_{2,1}|+e^{i\frac{2m\pi}{3}}|111\rangle\langle
S_{2,2}|)+\frac{\sqrt{\gamma}
\tilde{\delta}_{1}^{2}\Omega_{1}^{2}}{3\sqrt{6}\tilde{R}_{3,1}}|111\rangle\langle
S_{2,3}|,
\end{eqnarray}
\end{widetext}
where
\begin{eqnarray}\label{a7}
\tilde{\Delta}_{k}&=&\Delta_{k}-\frac{i\gamma}{2},
\tilde{\delta}_{k}=\Delta_{k}-\frac{i\kappa}{2}, \cr\cr
\tilde{R}_{n,k}&=&\tilde{\Delta}_{k}\tilde{\delta}_{k}-ng^{2},
\end{eqnarray}
and Re[ ] denotes the real part of the argument. The square of the
coefficient of each term of Eq. (\ref{a4}, \ref{a5}, \ref{a6})
describes the effective decay rate. For example,
$-\frac{\sqrt{\kappa}
g\Omega_{1}}{\sqrt{3}\tilde{R}_{1,1}}|S_{1,3}\rangle\langle000|$
indicates the decay from $|000\rangle$ to $|S_{1,3}\rangle$ at the
rate $|-\frac{\sqrt{\kappa}
g\Omega_{1}}{\sqrt{3}\tilde{R}_{1,1}}|^2$. We can obtain the
effective Hamiltonian and Lindblad operators of the $2$nd decay
process in the same way.

In the following, we study the $3$rd independent decay process
caused by the 3rd classical field, driving the transition
$|1\rangle\leftrightarrow|2\rangle$. The perturbation $V^{(3)}_{+}$
is written as
\begin{eqnarray}\label{a8}
V^{(3)}_{+}=\frac{\Omega_{3}}{3}\sum_{m=1}^{3}|2\rangle_{m}\langle1|.
\end{eqnarray}
Applying Eq. (\ref{e6}) and (\ref{e7}) to the process, we can obtain
the effective Hamiltonian and Lindblad operators
\begin{widetext}
\begin{eqnarray}\label{a9}
H_{eff,3}&=&\textrm{Re}\bigg[\frac{
\tilde{\delta}_{3}\Omega_{3}^{2}}{9\tilde{R}_{1,3}}\bigg](|S_{1,1}\rangle\langle
S_{1,1}|+|S_{1,2}\rangle\langle S_{1,2}|+|S_{1,3}\rangle\langle
S_{1,3}|)\cr&&+\textrm{Re}\bigg[\frac{2
\tilde{\delta}_{3}\Omega_{3}^{2}}{9\tilde{R}_{2,3}}\bigg](|S_{2,1}\rangle\langle
S_{2,1}|+|S_{2,2}\rangle\langle S_{2,2}|+|S_{2,3}\rangle\langle
S_{2,3}|)+\textrm{Re}\bigg[\frac{
\tilde{\delta}_{3}\Omega_{3}^{2}}{3\tilde{R}_{3,3}}\bigg]|111\rangle\langle
111|\\ \label{a10} L_{eff,3}^{\kappa}&=&\frac{\sqrt{\kappa}
g\Omega_{3}}{3\tilde{R}_{1,3}}(|S_{1,1}\rangle\langle
S_{1,1}|+|S_{1,2}\rangle\langle S_{1,2}|+|S_{1,3}\rangle\langle
S_{1,3}|) \cr&&+\frac{2\sqrt{\kappa}
g\Omega_{3}}{3\tilde{R}_{2,3}}(|S_{2,1}\rangle\langle
S_{2,1}|+|S_{2,2}\rangle\langle S_{2,2}|+|S_{2,3}\rangle\langle
S_{2,3}|)+\frac{\sqrt{\kappa}
g\Omega_{3}}{\tilde{R}_{3,3}}|111\rangle\langle111|,\\
\label{a11} L_{eff,3}^{\gamma,0,m}&=&\frac{\sqrt{\gamma}
\tilde{\delta}_{3}\Omega_{3}}{3\sqrt{6}\tilde{R}_{1,3}}(e^{i\frac{2m\pi}{3}}|000\rangle\langle
S_{1,1}|+e^{-i\frac{2m\pi}{3}}|000\rangle\langle
S_{1,2}|+|000\rangle\langle S_{1,3}|) \cr&&+\frac{\sqrt{\gamma}
\tilde{\delta}_{3}\Omega_{3}}{9\sqrt{2}\tilde{R}_{2,3}}(2e^{i\frac{2(m-2)\pi}{3}}|S_{1,1}\rangle\langle
S_{2,1}|-e^{i\frac{2\pi}{3}}|S_{1,2}\rangle\langle
S_{2,1}|-e^{-i\frac{2(m-1)\pi}{3}}|S_{1,3}\rangle\langle S_{2,1}|)
\cr&&+\frac{\sqrt{\gamma}
\tilde{\delta}_{3}\Omega_{3}}{9\sqrt{2}\tilde{R}_{2,3}}(-e^{-i\frac{2\pi}{3}}|S_{1,1}\rangle\langle
S_{2,2}|+2e^{-i\frac{2(m-2)\pi}{3}}|S_{1,2}\rangle\langle
S_{2,2}|-e^{i\frac{2(m-1)\pi}{3}}|S_{1,3}\rangle\langle S_{2,2}|)
\cr&&+\frac{\sqrt{\gamma}
\tilde{\delta}_{3}\Omega_{3}}{9\sqrt{2}\tilde{R}_{2,3}}(-e^{-i\frac{2m\pi}{3}}|S_{1,1}\rangle\langle
S_{2,3}|-e^{i\frac{2m\pi}{3}}|S_{1,2}\rangle\langle
S_{2,3}|+2|S_{1,3}\rangle\langle S_{2,3}|) \cr&&+\frac{\sqrt{\gamma}
\tilde{\delta}_{3}\Omega_{3}}{3\sqrt{6}\tilde{R}_{3,3}}(e^{i\frac{2(m-1)\pi}{3}}|S_{2,1}\rangle\langle111|
+e^{-i\frac{2(m-1)\pi}{3}}|S_{2,2}\rangle\langle111|+|S_{2,3}\rangle\langle111|),\\
\label{a12} L_{eff,3}^{\gamma,1,m}&=&\frac{\sqrt{\gamma}
\tilde{\delta}_{3}\Omega_{3}}{9\sqrt{2}\tilde{R}_{1,3}}(|S_{1,1}\rangle\langle
S_{1,1}|+e^{-i\frac{2m\pi}{3}}|S_{1,2}\rangle\langle
S_{1,1}|+e^{i\frac{2m\pi}{3}}|S_{1,3}\rangle\langle S_{1,1}|)
\cr&&+\frac{\sqrt{\gamma}
\tilde{\delta}_{3}\Omega_{3}}{9\sqrt{2}\tilde{R}_{1,3}}(e^{i\frac{2m\pi}{3}}|S_{1,1}\rangle\langle
S_{1,2}|+|S_{1,2}\rangle\langle
S_{1,2}|+e^{-i\frac{2m\pi}{3}}|S_{1,3}\rangle\langle
S_{1,2}|)\cr&&+\frac{\sqrt{\gamma}
\tilde{\delta}_{3}\Omega_{3}}{9\sqrt{2}\tilde{R}_{1,3}}(e^{-i\frac{2m\pi}{3}}|S_{1,1}\rangle\langle
S_{1,3}|+e^{i\frac{2m\pi}{3}}|S_{1,2}\rangle\langle
S_{1,3}|+|S_{1,3}\rangle\langle S_{1,3}|) \cr&&+\frac{\sqrt{\gamma}
\tilde{\delta}_{3}\Omega_{3}}{9\sqrt{2}\tilde{R}_{2,3}}(2|S_{2,1}\rangle\langle
S_{2,1}|-e^{i\frac{2(m-1)\pi}{3}}|S_{2,2}\rangle\langle
S_{2,1}|-e^{-i\frac{2(m-1)\pi}{3}}|S_{2,3}\rangle\langle S_{2,1}|)
\cr&&+\frac{\sqrt{\gamma}
\tilde{\delta}_{3}\Omega_{3}}{9\sqrt{2}\tilde{R}_{2,3}}(-e^{-i\frac{2(m-1)\pi}{3}}|S_{2,1}\rangle\langle
S_{2,2}|+2|S_{2,2}\rangle\langle
S_{2,2}|-e^{i\frac{2(m-1)\pi}{3}}|S_{2,3}\rangle\langle S_{2,2}|)
\cr&&+\frac{\sqrt{\gamma}
\tilde{\delta}_{3}\Omega_{3}}{9\sqrt{2}\tilde{R}_{2,3}}(-e^{i\frac{2(m-1)\pi}{3}}|S_{2,1}\rangle\langle
S_{2,3}|-e^{-i\frac{2(m-1)\pi}{3}}|S_{2,2}\rangle\langle
S_{2,3}|+2|S_{2,3}\rangle\langle S_{2,3}|) \cr&&+\frac{\sqrt{\gamma}
\tilde{\delta}_{3}\Omega_{3}}{3\sqrt{2}\tilde{R}_{3,3}}|111\rangle\langle
111|.
\end{eqnarray}
\end{widetext}
The effective Hamiltonian and lindblad operators of $4$th decay
process can be obtained in the same way.

\end{document}